\def\mna4{
 \hsize=6.3in
 \vsize=9.6in
 \voffset=-0.3in
}

 \font\twelvebf=cmbx12
 \font\twelvett=cmtt12 at 12pt
 \font\twelverm=cmr12 at 12pt    \font\ninerm=cmr9
\font\sevenrm=cmr7    \font\fiverm=cmr5
 \font\twelvei=cmmi12 at 12pt    \font\ninei=cmmi9
\font\seveni=cmmi7    \font\fivei=cmmi5
 \font\twelvesy=cmsy10 at 12pt   \font\ninesy=cmsy9
\font\sevensy=cmsy7   \font\fivesy=cmsy5
 \font\twelveit=cmti12 at 12pt
 \font\twelvesl=cmsl12 at 12pt
 \font\tenex=cmex10
 \font\teni=cmmi10    \font\tensy=cmsy10
 \font\tentt=cmtt10   \font\tenit=cmti10
 \font\tensl=cmsl10

\def\mntwelvepoint{\def\rm{\fam0\twelverm}
 \textfont0=\twelverm \scriptfont0=\ninerm \scriptscriptfont0=\sevenrm
 \textfont1=\twelvei \scriptfont1=\ninei \scriptscriptfont1=\seveni
 \textfont2=\twelvesy \scriptfont2=\ninesy \scriptscriptfont2=\sevensy
 \textfont3=\tenex \scriptfont3=\tenex \scriptscriptfont3=\tenex
 \def\it{\fam\itfam\twelveit}%
 \textfont\itfam=\twelveit
 \def\sl{\fam\slfam\twelvesl}%
 \textfont\slfam=\twelvesl
 \def\bf{\fam\bffam\twelvebf}%
 \textfont\bffam=\twelvebf
 \def\tt{\fam\ttfam\twelvett} 
 \textfont\ttfam=\twelvett
 \baselineskip 14pt%
 \abovedisplayskip 14pt plus 3pt minus 10pt%
 \belowdisplayskip 14pt plus 3pt minus 10pt%
 \abovedisplayshortskip 0pt plus 3pt%
 \belowdisplayshortskip 8pt plus 3pt minus 5pt%
 \parskip 3pt plus 1.5pt
 \setbox\strutbox=\hbox{\vrule height10pt depth4pt width0pt}%
 \rm}
\def\mntenpoint{\def\rm{\fam0\tenrm}
 \textfont0=\tenrm \scriptfont0=\sevenrm \scriptscriptfont0=\fiverm
 \textfont1=\teni \scriptfont1=\seveni \scriptscriptfont1=\fivei
 \textfont2=\tensy \scriptfont2=\sevensy \scriptscriptfont2=\fivesy
 \textfont3=\tenex \scriptfont3=\tenex \scriptscriptfont3=\tenex
 \def\sl{\fam\itfam\tenit}%
 \textfont\itfam=\tenit
 \def\sl{\fam\slfam\tensl}%
 \textfont\slfam=\tensl
 \def\bf{\fam\bffam\tenbf}%
 \textfont\bffam=\tenbf
 \def\tt{\fam\ttfam\tentt}%
 \textfont\ttfam=\tentt
 \baselineskip 12pt%
 \abovedisplayskip 14pt plus 3pt minus 10pt%
 \belowdisplayskip 14pt plus 3pt minus 10pt%
 \abovedisplayshortskip 0pt plus 3pt%
 \belowdisplayshortskip 8pt plus 3pt minus 5pt%
 \parskip 3pt plus 1.5pt
 \setbox\strutbox=\hbox{\vrule height8.5pt depth3.5pt width0pt}%
 \rm}
\documentstyle {article}
\begin{document}
\title{Are Disappearing Dwarfs Just Lying Low?}
\author{Steven Phillipps$^{1}$ and Simon Driver$^{2}$\\
$^{1}$ Department of Physics,
University of Bristol\\ Tyndall Avenue,
Bristol BS8 1TL\\
$^{2}$ Department of Physics and Astronomy, Arizona State University,\\
Box 871504, Tempe, Arizona, USA}
\maketitle

\begin{abstract}
Recent redshift surveys have shown that the excess galaxies seen in faint
galaxy number counts (above those expected given the local
galaxy luminosity function) are not evolved giants at high redshifts, but
low to moderate luminosity objects at more modest redshifts. This has led to
the suggestion that there was once an additional population of dwarf galaxies
which has since disappeared, ie. there is non-conservation of galaxy number.
Here we investigate the possibility that these disappearing dwarfs
have actually evolved to become the population of very low surface brightness
galaxies which is now being detected in nearby clusters.

\end{abstract}

{\bf Key words:} galaxies:photometry - galaxies:luminosity function
- galaxies:evolution

\section{The Problem}
It has been known for many years (eg. Kron 1978, Peterson et al 1979) that
at faint
magnitudes the number-magnitude counts of galaxies exceed the expectations
for a non-evolving galaxy population, based on the locally observed luminosity
function.
More recently redshift surveys (Broadhurst, Ellis \& Shanks 1988 =
BES; Colless et al 1990 = CETH; Cowie, Songaila \& Hu 1991 = CSH; Colless et
al 1993) have shown that the excess (generally blue)
galaxies are not very distant evolved giants which have been brightened
sufficiently to enter the surveys, but relatively nearby dwarfs, thus
requiring high volume densities of galaxies. Indeed, at the
extreme limits reached in the counts in recent years (eg. Tyson 1988; Metcalfe
et al 1991 = MSFJ; Metcalfe, Shanks \& Fong 1992), it has been claimed that
there is just not enough total volume in a
standard (ie. zero cosmological constant) universe out to $z \simeq 3$ (where
the Lyman continuum break
enters the optical bands) to hold all the observed galaxies,
whatever their distances and brightnesses $if$ they have the same volume
density as conventionally assumed for present day galaxies (Koo 1989;
Fukugita et al 1990).

It has therefore been suggested by CSH and others (eg. Babul \& Rees 1992)
that at moderate redshifts (around 0.5, say) there must have been a large
additional population of dwarf galaxies which is no longer present. The
suggestions summarised by Lacey (1991) and Guiderdoni (1993)
are that these dwarfs have since merged to form
today's giant galaxies, that they have self-destructed as a result of the
supernova explosions after their active star forming phase, or that they have
faded so much that they are now invisible. Here we investigate in more detail
a case (discussed briefly in Phillipps 1993a) which is akin to the third of
these, namely that the medium redshift dwarfs have faded
to become the present day population of low surface brightness galaxies
(LSBGs) which are turning up in increasing numbers and at increasingly faint
surface brightnesses in recent surveys (eg. Binggeli, Sandage \& Tammann 1985;
Impey, Bothun \& Malin 1988 = IBM; Irwin et al 1990 =
IDDP; Schombert et al 1992; Turner et al 1993 = TPDD).

\section{Disappearing Dwarfs}

\subsection{Numbers}

Consider first how many dwarfs we require.
According to the counts reviewed by CSH, the surface density of
galaxies to a B magnitude limit of 26.5 is about $2.5 \times 10^{5}$ per square
degree. Now the corresponding total volume out to $z=3$ is $2.2 \times 10^{7}$
Mpc$^{3}$ for an $\Omega = 1$ universe with $H_{0} = 50$ km s$^{-1}$
Mpc$^{-1}$ (ie. $h=0.5$ in the usual notation). Thus the $minimum$ volume
density of $observable$ galaxies
required is about 0.012 per Mpc$^{3}$. This is then to be
contrasted with the canonical value of about 0.0006 for galaxies brighter than
$L_{*}$ (we assume the characteristic magnitude for giants to be about
$M_{*} = -21.3$ in $B$, see below). If instead we concentrate on the galaxies
seen at slightly
brighter magnitudes which are (just) accessible to redshift surveys, at $B =
24$, then there are about $2 \times 10^{4}$ galaxies per square degree. CSH
note that half of these lie at $z < 0.5$. Using the volume out to $z=0.5$
we get a mean observable volume density
over this region of about $10^{4}/10^{6} \simeq 0.01$. Note that
the limiting magnitude $B = 24$ corresponds, at $z=0.5$, to an absolute
magnitude $M_{B} =
-19.0 - K_{2}(0.5)$, where $K_{2}(z)$ is the spectrum
dependent part of the k-correction (ie. excluding the bandwidth term). We
therefore need a density of approximately 0.01 galaxies per
Mpc$^{3}$ brighter than $M_{B} = -19$ (ie. around $M_{*} + 2.3$) even
if $K_{2}(0.5) \simeq 0$. Of
course many of the galaxies are nearer than $z=0.5$ so some absolutely fainter
galaxies will also be included in the counts,
while using a low $\Omega$ model can also alleviate the problem (though only
slightly for redshifts less than 0.5).
Hence this is a worst case analysis
in terms of trying to find enough dwarfs to fit the counts.

If we consider first a 'standard' local luminosity function (eg. Phillipps  \&
Shanks 1987; Efstathiou, Ellis \& Peterson 1988; Loveday et al 1992 = LPEM),
viz. a
Schechter (1976) function with $\phi_{*} = 0.0022$, $M_{*} = -21.3$, $\alpha =
-1.1$ (again assuming $h = 0.5$), then we only reach the
necessary density of 0.01 galaxies per cubic Mpc at $M \simeq M_{*} + 6.5
\simeq -14.8$. Thus, if we assume a roughly flat spectrum for a star forming
dwarf (eg. Cowie et al 1988)
(ie. no spectral k-correction term), then we require over
4 magnitudes of brightening by $z= 0.5$
\footnote{Notice that this rate of evolution would be
sufficient for almost all the
galaxies visible at say $z \sim 0.1$ to $still$ be visible at $z \sim 0.5$, ie.
the evolutionary brightening almost completely counteracts the increasing
distance modulus. The observable galaxy density would therefore remain more or
less fixed and the 'worst case' assumption would be close to the truth.}
. For a higher local normalisation this could be
lowered slightly, but even so this simple calculation confirms the view (as
in Lacey's (1991)
review) that it is unlikely that the dwarfs could have faded enough
over the available time interval for this scheme to work, in the standard
picture (see also Guiderdoni \& Rocca-Volmerange 1991, Lacey et al 1993).

However, at present-day magnitudes around $-15$ many galaxies are of low
(or very low) surface brightness and are therefore missing from conventional
luminosity functions.
(That is, they fail to meet the criteria for inclusion in a sample based on
isophotal magnitude or diameter: see Sandage, Binggeli \& Tammann (1985) for
a discusson of a famous historical instance of this effect).
It now appears (Sandage et al 1985;
Phillipps et al 1987; IBM; Impey \& Bothun 1989; IDDP; Ferguson 1990)
that, at least in clusters,
allowing for these LSBGs steepens the faint end of the
luminosity function considerably.
Schade \& Ferguson (1994) suggest that this may be true in the field too.
(See also Marzke et al (1994) and Marr (1994) for field galaxy LFs with
evidence for a turn up at the faint end).
At magnitudes fainter than about $-17.5$, the
effective value of $\alpha$ is about -1.5 or even steeper.
Driver et al (1994a = DPDMD) discuss in more detail the question of
such steep faint
end slopes, which are theoretically expected on quite general grounds (see eg.
White \& Frenk 1991; Lacey et al 1993; Kauffmann, White \& Guiderdoni 1993).
Note particularly that in such models the overall
LF no longer has a simple single Schechter function form (Binggeli, Sandage \&
Tammann 1988; Ferguson \& Sandage 1991; Driver et al 1994b). In this
case, even with a relatively low normalisation for the bright end
($\phi_{*} = 0.0018$) we get 0.01 galaxies
per Mpc$^{3}$ brighter than about $M_{B} = -16.5$ (depending on the exact point
where the steep part of the LF cuts in; again see DPDMD for a
discussion). This then leads to a much more reasonable requirement of $\leq
2.5^{m}$ of brightening at $z = 0.5$. Recall that we have deliberately chosen
worst cases throughout, so the actual evolution required may be
less than this.
Indeed, DPDMD considered the option of fitting the number counts
using an extremely steep
faint end slope for the present-day LF and $no$ evolution (see also Guiderdoni
1992; Koo \& Kron 1992; Koo, Gronwall \& Bruzual 1993), though this may run
into the problem of seriously
$under$predicting the mean redshift at faint magnitudes.

The most recent redshift surveys do now appear to show direct evidence for a
change in the LF shape with $z$ (Ellis 1994; Colless
1994), thus
indicating some sort of mass dependent evolutionary process along the general
lines originally suggested by BES (see also Cole, Treyer \& Silk 1992).
We therefore wish to consider as a next simplest model (after no evolution and
general luminosity evolution), the case where we have only simple fading of the
(steep) dwarf component of the LF.

\subsection{Fading}

Next consider the levels of fading which might be reasonable for a dwarf galaxy
between $z=0.5$ and the present. If the dwarfs are seen in a star-forming dI
like state (eg. Lin \& Faber 1983, Tyson \& Scalo 1988,
Davies \& Phillipps 1988) they can
be expected to fade by about $2.5^{m}$ over the $6 \times 10^{9}$
years look-back time since $z = 0.5$ according to models of simple ageing of
starbursts (eg. Wyse 1985; Guiderdoni \& Rocca-Volmerange 1987; Charlot \&
Bruzual 1991). This rate would then be sufficient to produce even
the 'worst case' evolution considered above.

If we assume that the fading occurs at a fixed galaxy size (eg. Bothun et al
1986, Evans et al 1990; though see also Colless et al 1994),
the decay in surface brightness is exactly the same as that in total
luminosity. Thus galaxies could have had central surface brightnesses around
22.5 $B \mu$ (ie. $B$ magnitudes per square arc second)
as dIs and have faded to become LSBGs with central surface brightnesses
around 25 $B \mu$. These would be just barely detectable in $dedicated$ local
LSBG surveys (eg. IBM, IDDP). Anything with a dI
surface brightness (at $z=0.5$) at or below 24 $B \mu$ (which is apparently
quite common, see eg. TPDD) would by now be quite invisible locally
(unless it was a satellite of our Galaxy).

It is worth remarking as a general
point (Phillipps 1993a; see also McGaugh 1994; Lacey et al 1993)
that while distant galaxies are frequently surveyed with
extremely low isophotal thresholds around 29 $B \mu$ (eg. Tyson 1988;
Metcalfe et al 1992; Tresse et al 1993), corresponding to an intrinsic surface
brightness $\sim 27.5 B \mu$ allowing for cosmological dimming, virtually
nothing is known about $local$ galaxies at surface brightness
levels below about 25 $B \mu$, the deliberate LSBG searches in the Fornax and
Virgo Clusters (eg. IBM, IDDP) being the only real exceptions.
This paradoxical situation can obviously have
important implications when trying to interpret the faint counts in terms of
evolutionary rates. Clearly a
galaxy with an intrinsic central surface brightness around 26.5 $B \mu$ which
should be
readily observable in the deep counts (its observed surface brightness will be
about $1^{m}$ above the limiting threshold), with $any$ evolutionary fading at
all will not be included in $any$ local census. This loss of galaxies nearby
because of their low surface brightness as well as magnitude may therefore
further reduce the amount of evolution actually needed to generate the apparent
increase in numbers at fainter magnitudes/larger redshifts. We explore this
more quantitatively in section 3.2.

\section{Fading Dwarf Models}

\subsection{Luminosity Functions and Evolution}

To make a more rigorous assessment of the viability of the simple fading
dwarf picture, we construct a model with a two component LF as advocated in
DPDMD. The first component is a standard Schechter function for the
giants with $\alpha = -1$ and $M_{*} = -21$, but cut off at the faint end (cf.
Sandage et al 1985). For added simplicity we take
this to be non-evolving.
The lack of evolved very distant giants already limits the allowed
evolution to be less than 1 magnitude by $z = 1$ (Colless 1993). The
normalization $\phi_{*}$(giant) is chosen so as to obtain a fit to the counts
at $B = 18$ to 20. The second,
dwarf, component has the steep slope ($\alpha = -1.5$) discussed earlier,
a present day $M_{*} = -18$ (ie. 3 magnitudes or a factor 16 fainter than
for giants) and an
amplitude relative to the giant LF which we leave as a free parameter. This is
equivalent to being able to choose the absolute magnitude at which the dwarfs
begin to dominate the overall LF (see DPDMD).

We now want to allow this population to fade at a rate consistent with
that expected
after a starburst of duration, say, 1 Gyr (a reasonable liftime for the star
forming phase of a dwarf galaxy, see Davies \& Phillipps 1987).
As a straightforward,
easily parameterised form we model this as an exponential fall off in
luminosity with time (see also Cole et al 1992), as commonly adopted for the
variation of the star formation rate (eg. Guiderdoni \& Rocca-Volmerange 1987).
This then translates to an evolutionary correction (in magnitudes) which is
linear with lookback time, ie. $\Delta m \propto \Delta t$.
Since for an Einstein-de Sitter universe,
$t \propto (1+z)^{-3/2}$, for convenience we can write the evolutionary
correction as  $\Delta m = 4 \beta \Delta t/t_{0} =
4 \beta [1 - (1+z)^{-3/2}]$. Note that if $\beta = 1$ the fading since $z=0.5$
is 1.8 magnitudes, while if $\beta = 1.5$ this rises to 2.7 magnitudes, so
this would be the range that one might expect for dwarfs fading after a burst
of star formation (cf. Wyse (1985), Guiderdoni \& Rocca-Volmerange (1987) and,
especially, Olofsson (1989) who considers low metallicity systems).
Note, too, that this form of the evolution avoids excessive values at high $z$
(cf. Phillipps 1993b) since the evolution soon flattens out beyond
$z \sim 0.5$.
In terms of the number of dwarfs we would see, the fact that only the power
law tail of the dwarf LF is important means that, say,  $2^{m}$ of fading
is essentially equivalent to decreasing the normalization of the
dwarf LF by a factor $\Phi(M+2)/\Phi(M) = 10^{0.4 (-\alpha -1) \times 2}
= 2.5$.

To see if such a simple model can fit the observations, we take as our key
constraints (i) the faint end of the local LF, (ii) the shape of the $B$ band
number counts faintwards of $B = 20$
\footnote{Even in an extreme dwarf dominated model,
such as in DPDMD, the counts brighter than this are mostly due to the giants.
Fits to this part and possible evolutionary effects on it are discussed by
Maddox et al (1990), Eales (1993) and Lonsdale \& Chokshi (1993).}
and (iii) the redshift distributions at $B \simeq 21$, 22 and 23.5. We could,
of course, attempt to introduce further constraints by using, in particular,
the $K$ band number counts (eg. Gardner, Cowie \& Wainscoat 1993).
However, the number counts at $K$ are
much less steep than at $B$, so present less of a 'problem'. Less evolution
and/or less dwarfs are needed and would, indeed, be expected from the evolution
models (see eg. Phillipps 1993b) and the fact that the dwarfs are generally
blue so contribute relatively little to counts at longer wavelengths
(eg. DPDMD).

If for the moment we ignore the (un)detectability of low surface brightness
objects (ie. we consider only total magnitudes), then it is easy to generate
these various distributions. Figures 1a-e show the predictions for a
basic model with $\beta=1$ and a
normalization $\Gamma = \phi_{*}$(dwarf)/$\phi_{*}$(giant) = 1,
chosen to give a respectable overall fit compared to the recent
data (from LPEM, MSFJ, BES, CETH and
CSH, respectively). Notice that the redshift distributions, in particular,
follow the observations quite closely.
The distribution at $B = 23$ to 24 is noteable for its very large width;
galaxies are almost equally likely to be at redshifts anywhere between 0.1 and
0.6. This is generally true for the other models, below, too.
The local LF is perhaps not quite as
flat as LPEM's best fit $model$ in the range $-18$ to $-16.5$ but it is
nevertheless a perfectly good fit to the data points. However,
it can be seen that the count slope is still
not sufficiently steep at $B > 23$ (though remember we are
using $\Omega = 1$ which does make it harder to match the slope).

We can now alter the evolutionary rate and the present
dwarf to giant ratio to try to improve these fits. If we set $\Gamma = 2$,
ie. double the number of dwarfs, we obviously get higher counts at faint
magnitudes, and from figure 2b we can see that the fit to the counts is now
very good (compared to the scatter in the data points). The redshift
distributions (not shown) are still in fairly satisfactory agreement with
observation, though
rather more low $z$ objects are predicted at $B = 21$ than are actually seen.
However the fit to the LPEM LF is much poorer, with significantly too many
galaxies at $M_{B} = -18$ to $-16.5$ (see figure 2a).

Another potentially successful alternative is to increase the evolutionary
rate. If we set $\beta = 1.5$ (keeping $\Gamma = 1$) then we again obtain an
excellent
fit to the number counts (even to the slight inflection at $B = 21$ and the
curvature around $B = 24$; see figure 3b). The local LF remains the same as
before (figure 1a), but now the redshift distributions at $B = 22$
and 23.5
develop tails to large $z$ which are not seen in the data (figures 3d and 3e).
Statistics of the gravitational lensing of faint background objects by clusters
at different redshifts suggest that this lack of high $z$ objects may also
extend to even fainter magnitudes (see Smail, Ellis \& Fitchett 1994).
We might try to remove this
problem by cutting the evolution off at $z \simeq 0.7$, eg. by assuming a peak
in the star formation at that epoch, but this destroys the
previously excellent fit to the faint counts (figure 4b)
since an increasing fraction of the galaxies at $B
> 23$ were at high $z$ in this model (eg. figure 3e).

Metcalfe et al (1994a,b) have emphasised that if there is a redshift cut-off
(or more generally if there is no significant tail to high $z$) then
the slope of the very faint counts should reflect the slope of the LF. This
suggests as another alternative a dwarf LF with $\alpha = - 1.75$, similar to
those in DPDMD, Guiderdoni (1991) and Lacey et al (1993).
This does indeed provide an excellent fit to the faint counts (figure 5b) but
even after fading, there are still too many local dwarfs (figure 5a), so this
model shares the same problems as the non-evolving dwarf dominated model of
DPDMD.

Thus it appears that none of the models which give good fits to the number
counts are simultaneously good fits to the redshift survey data and/or the
local LF. However, there remains the question of surface brightness bias in the
nearby data, as mentioned at the end of section 2.

\subsection{Surface Brightness Effects}

For more realistic modelling, then, we should also allow for the loss
of galaxies due to low
surface brightness (cf. Phillipps, Davies \& Disney 1990; Lacey et al 1993;
McGaugh 1994).
For simplicity we will assume that galaxies are detectable if their
central surface brightnesses are above some threshold (as in eg. Tresse et al
1993). In practise most surveys would also require that the isophotal size,
magnitude and/or signal to noise ratio exceeded some threshold too (see eg.
Phillipps \& Disney 1986; IDDP; Yoshii 1993; TPDD; DPDMD) but we ignore this
added
complication here. (It is dealt with in more detail in papers by Bristow
\& Phillipps and Ferguson \& McGaugh currently in preparation).
We also ignore the effects of atmospheric seeing (eg. Ellis, Fong \& Phillipps
1977; Yoshii 1993).

To make the surface brightness corrections
we need to assume some form for the bivariate
distribution of galaxy central surface brightness (or intensity $I$) and scale
size $a$ (see Phillipps et al 1990). If we take
the relations discussed in IDDP (see also Bothun, Impey \& Malin 1991), viz.
$n(a,I) da \, dI \propto a^{-2} I^{-1} da \, dI$ for
$I \leq I_{max}$ (which self-consistently
gives an $L^{-3/2}$ luminosity function), then we can easily calculate the
fraction $f$ of the galaxies at any $L$ which are also brighter than some
surface brightness limit $I_{min}$.
Although the total numbers clearly increase towards low surface intensities
(ie. there are many LSBGs), the preponderance of small galaxies means that {\em
at fixed L} most of the galaxies have small $a$ but values of $I$ towards the
top of its range. This is therefore by no means an extreme model, as far as
invoking surface brightness selection is concerned.
In fact, since $a^{2} \propto L/I$, we will have
\[n(L,I) dL \, dI = n(a,I) (\partial a / \partial L)_{I} \, dL \, dI \]
\[\Rightarrow n(L;>I_{min}) dL \propto \int_{I_{min}}^{I_{max}} I^{-1}
(I/L) I^{-1/2} L^{-1/2} dL \, dI \]
\[\propto L^{-3/2} \, (I_{max}^{1/2} - I_{min}^{1/2}). \]
Clearly the total LF is just given by this expression with $I_{min} = 0$,
so the fraction of galaxies above any $I_{min}$ is simply
\[f = 1 - (I_{min}/I_{max})^{1/2}, \]
independent of $L$. Thus the effect
is equivalent to an overall renormalisation of the dwarf LF by the factor $f$.
Of course, this is only an approximation to the actual observational situation,
where, as noted above, other factors besides the central surface brightness
itself are likely to influence the selection for a sample. (Essentially it
amounts to assuming that the selection boundary is a horizontal line in the
$(L, I)$ plane, rather than a somewhat curved one, as in eg. TPDD's  figure
3, where the effective limiting $I_{min}$ changes by about 0.5 $B \mu$ across
a 4 magnitude range in $L$).

Since the
observed $I_{max}$ depends on $z$ because of evolution, cosmological dimming
etc., this could  mimic an $extra$ evolutionary correction within a given
survey (for instance, less galaxies might pass the surface brightness test
at $z = 0$ if
they have faded substantially). However, it turns out that with realistic
fading rates, as above,
the evolutionary changes to the surface brightness more or less
cancel with the cosmological dimming, so in any one survey (ie. for fixed
$I_{min}$) there will be no significant effect (since $I_{max}$ is also
approximately fixed). The more important effects therefore arise $between$
surveys, since generally deep surveys have a much fainter limiting isophote
than local surveys (Phillipps 1993a; McGaugh 1994; Schade \& Ferguson 1994).
For instance, if we take
$I_{max}$ to correspond to $22.5 B\mu$ (IDDP) then for surveys with isophotal
thresholds around $24.5 B\mu$ (appropriate for most local luminosity function
determinations),
$25.5 B\mu$ (for moderately faint redshift surveys), $26.5 B\mu$ (deep redshift
surveys) and $29 B\mu$ (very faint counts), we get values of $f$ of,
respectively, 0.60, 0.75, 0.84 and 0.95.
\footnote{If we require the central surface brightness to exceed the limiting
isophote by a at least 0.5 $B \mu$ (a reasonable value in practice, see eg.
TPDD) then the figures for $f$ reduce further to 0.50, 0.68, 0.80 and 0.94.
Conversely, if we increased the bright limit to correspond to, say,
21.5 $B \mu$ (eg. van der Kruit 1987, Phillipps et al 1987)
the $f$s would increase again. However, we would class these objects with our
'normal' galaxy population.}
Incorporating these corrections
clearly enables us to have more dwarfs in the deeper data without affecting the
local LF. Figures 6a-e show such a model, with $\beta = 1$, $\Gamma = 2$, ie.
the 'standard' amount of evolution and a high true number of dwarfs.
As expected this leads to a better
allround fit than the previous models since in terms of the counts we
essentially have the $\Gamma = 2$ model from above, but for the local LF the
{\it effective} $\Gamma$ is only 1.2, that is almost half the dwarfs have
too low a
surface brightness to be included. A very similar fit to the counts
is achieved if we allow
slightly more evolution (eg. $\beta = 1.25$) and reduce the local dwarf numbers
somewhat ($\Gamma = 1.75$), thus improving the (already acceptable) fit to the
LPEM data. Figure 7 shows how the LF would look at different redshifts under
our 'best' model of figure 6. This might be compared with recent observations
reported by Eales (1993), Colless (1994) and Ellis (1994) or Lilly (1994) who
variously claim evolution in $\phi_{*}$, $\alpha$ or $M_{*}$.

\section{Discussion}

Several authors (eg. Ostriker 1990; Delcanton 1993; Roukema \& Yoshii 1993)
have recently outlined reasons why the otherwise attractive merger picture (eg.
Rocca-Volmerange \& Guiderdoni 1990; Broadhurst, Ellis \& Glazebrook 1992;
Carlberg 1992) may
not be the answer to the large numbers of dwarfs seen at moderate redshifts but
not today. It is therefore still important to consider models
which do conserve galaxy numbers. As non-evolving standard (ie. giant
dominated) models, evolving giant models and non-evolving dwarf dominated
models all appear to be ruled out (the first from the numbers, the other two
from the redshift distributions), the next simplest model to try is probably
the dwarf dominated model with uniform fading of the dwarf component. This is
somewhat simpler than the original BES model which has progressively more
evolution for fainter and fainter systems or those of Cole et al (1992) where
the LF slope evolves directly. Although physically based on the
possible evolution of star forming dwarfs, it is still a fairly
phenomenological model, rather than a completely $ab$ $initio$ physical model
such as those attempted by Lacey et al (1993) or Kauffmann et al (1993),
for instance. Nevertheless it shares, in practice, many of the attributes of
the Kauffmann, Guiderdoni \& White (1994) model based on the hierarchical
merger of CDM halos, wherein dwarf galaxies form most of their stars prior to
the merger of their halos with those of larger galaxies and then
fade at a rate equivalent to that assumed above.

Our results suggest that fading of a large dwarf population - galaxies going
from star forming bright irregulars to quiescent
low surface brightness dwarfs - can
indeed account for the various observational constraints in a consistent way,
even if $\Omega = 1$. The required rate for the fading ($\beta
\simeq 1 - 1.25$) corresponds to 1.8 to 2.2 magnitudes since $z = 0.5$,
in excellent
agreement with theoretical modelling of post starburst evolution of stellar
populations. The corresponding e-folding time for the luminosity is close to 3
Gyr (for our assumed cosmology). In general, it need not be the case that the
evolution of the LF follows that of the individual galaxies. However, if the
galaxies share an evolutionary e-folding time scale then this will in fact
be so, at least after the point where most of them have started fading. Thus
for the current approximation to be valid, all we really require is that many
dwarfs had already experienced their major star forming episodes and had begun
to fade by $\sim 6$ Gyr ago. This would again be in line with the hierarchical
models aluded to above, where small halos suffer mergers with larger systems
at fairly early epochs.

In order to simultaneously fit the steep number count slope yet have no high
redshift galaxies requires a large density of dwarfs (since the evolution can
not be too strong). In order to reconcile this with the brighter redshift
surveys and the local LF then requires a model with a significant bias against
nearby dwarfs due to their low surface brightness. A simple calculation based
on the bivariate brightness distribution seen for dwarfs in the Fornax cluster
provides the basis for a quantitative assessment of this effect and shows
that it is indeed sufficient to remove the remaining discrepancy. If this model
is correct then we should expect to see a turn up in the field LF soon after
the last well determined point of current surveys and to find significant
numbers of low surface brightness field galaxies. Furthermore, we would
predict that when redshift surveys are able to probe even deeper than currently
possible we should continue to see primarily star forming (irregular ?) low
mass systems at a wide range of redshifts.

\section*{Acknowledgements}

We would like to thank Tom Broadhurst, Matthew Colless, Jon Davies,
Richard Ellis, Harry Ferguson, John Huchra, Dave
Koo, Stacy McGaugh and Nigel Metcalfe for useful discusssions.
We thank the referee for detailed and constructive comments on the first
version of the paper. SP acknowledges
the support of the Royal Society via a University Research Fellowship. SPD
thanks the University of Wales College of Cardiff for financial
support during the early part of this work.

\section*{References}

Babul A., Rees M.J., 1992, MNRAS, 255, 346

\noindent Binggeli B., Sandage A., Tammann G.A., 1985, AJ, 90, 1681

\noindent Binggeli B., Sandage A., Tammann G.A., 1988, ARA\&A, 26, 509

\noindent Bothun G.D., Impey C.D., Malin D.F., 1991, ApJ, 376, 404

\noindent Bothun G.D., Mould J.R., Caldwell N., MacGillivray H.T., 1986, AJ,
92, 1007

\noindent Broadhurst T.J., Ellis R.S., Glazebrook K.G., 1992, Nature, 355, 55

\noindent Broadhurst T.J., Ellis R.S., Shanks T., 1988, MNRAS, 235, 827

\noindent Carlberg R.G., 1992, ApJ, 399, L31

\noindent Charlot S., Bruzual G., 1991, ApJ, 367, 126

\noindent Cole S.M., Treyer M.A., Silk J., 1992, ApJ, 385, 9

\noindent Colless M.M., 1993, in First Light in the Universe: Stars or QSOs?,
eds Rocca-Volmerange B. et al

(Editions Frontieres)

\noindent Colless M., 1994, in Proceedings of the 35th Herstmonceux Conference,
Wide Field Spectroscopy and the

Distant Universe, in press

\noindent Colless M., Ellis R.S., Taylor K., Hook R.N., 1990, MNRAS, 244, 408

\noindent Colless M., Ellis R.S., Broadhurst T.J., Taylor K., Peterson B.A.,
1993, MNRAS, 261, 19

\noindent Colless M., Schade D., Broadhurst T.J., Ellis R.S., 1994, MNRAS, 267,
1108

\noindent Cowie L.L., Lilly S.J., Gardner J., McLean I.S., 1988, ApJ, 332, L29

\noindent Cowie L.L., Songaila A., Hu E.M., 1991, Nature, 354, 400

\noindent Davies J.I., Phillipps S., 1988, MNRAS, 233, 533

\noindent Delcanton J.J., 1993, ApJ, 415, L87

\noindent Driver S.P., Phillipps S., Davies J.I., Morgan I., Disney M.J.,
1994a, MNRAS, 266, 155

\noindent Driver S.P., Phillipps S., Davies J.I., Morgan I., Disney M.J.,
1994b, MNRAS, 268, 393

\noindent Eales S., 1993, ApJ, 404, 51

\noindent Efstathiou G., Ellis R.S., Peterson B.A., 1988, MNRAS, 232, 431

\noindent Ellis R.S., 1994, in Proceeedings of IAU Symposium No. 164, in press

\noindent Ellis R.S., Fong R., Phillipps S., 1977, MNRAS, 181, 163

\noindent Evans Rh., Davies J.I., Phillipps S., 1990, MNRAS, 245, 164

\noindent Ferguson H.C., 1990, Ph.D. Thesis, John Hopkins University

\noindent Ferguson H.C., Sandage A., 1991, AJ, 101, 765

\noindent Fukugita M., Takahara F., Yamashita K., Yoshii Y., 1990, ApJ, 361, L1

\noindent Gardner J.P., Cowie L.L., Wainscoat R.J., 1993, ApJ, 415, L9

\noindent Guiderdoni B., 1992, in The Early Observable Universe from Diffuse
Backgrounds, eds. Rocca-Volmerange

B. et al (Editions Frontieres), p.193

\noindent Guiderdoni B., 1993, in First Light in the Universe: Stars or QSOs?,
eds Rocca-Volmerange B. et al,

(Editions Frontieres)

\noindent Guiderdoni B., Rocca-Volmerange B., 1987, A\&A, 186, 1

\noindent Guiderdoni B., Rocca-Volmerange B., 1991, A\&A, 252, 435

\noindent Impey C.D., Bothun G.D., 1989, ApJ, 341, 89

\noindent Impey C.D., Bothun G.D., Malin D.F., 1988, ApJ, 330, 634

\noindent Irwin M.J., Davies J.I., Disney M.J., Phillipps S., 1990, MNRAS, 245,
289

\noindent Kauffmann G., White S.D.M., Guiderdoni B., 1993, MNRAS, 264, 201

\noindent Kauffmann G., Guiderdoni B., White S.D.M., 1994, MNRAS, 267, 981

\noindent Koo D.C., 1989, in The Epoch of Galaxy
Formation, eds. C.S. Frenk et al (Kluwer, Dordrecht), p.71

\noindent Koo D.C., Gronwall C., Bruzual G.A., 1993, ApJ, 415, L21

\noindent Koo D.C., Kron R.G., 1992, ARA\&A, 30, 610

\noindent Kron R.G., 1978, Ph.D. Thesis, University of California at Berkeley

\noindent Lacey C.G., 1991, Nature, 354, 430

\noindent Lacey C.G., Guiderdoni B., Rocca-Volmerange B., Silk J., 1993, ApJ,
402, 15

\noindent Lilly S.J., 1994, in Proceedings of the 35th Herstmonceux Conference,
Wide Field Spectroscopy and the

Distant Universe, in press

\noindent Lin D.C., Faber S.M., 1983, ApJ, 226, L21

\noindent Lonsdale C.J., Chokshi A., 1993, AJ, 105, 1333

\noindent Loveday J., Peterson B.A., Efstathiou G., Maddox S.J., 1992, ApJ,
390, 338

\noindent Maddox S.J, Sutherland W.J., Efstathiou G., Loveday J., Peterson
B.A., 1990, MNRAS, 247, 1P

\noindent Marr J.H., 1994, M.Sc. Thesis, University of Durham

\noindent Marzke R.O, Geller M.J., Huchra J.P., Corwin H.G., 1994, ApJ, 428, 43

\noindent McGaugh S.S., 1994, Nature, 367, 538

\noindent Metcalfe N., Shanks T., Fong R., Jones L., 1991, MNRAS, 249, 498

\noindent Metcalfe N., Shanks T., Fong R., 1992, Gemini, 34, 12

\noindent Metcalfe N., Shanks T., Roche N., Fong R., 1994a, in Astronomy from
Wide-Field Imaging, IAU

Symposium No. 161, ed. MacGillivray H.T. et al, p.645

\noindent Metcalfe N., Shanks T., Roche N., Fong R., 1994b, MNRAS, in press

\noindent Olofsson K., 1989, A\&AS, 80, 317

\noindent Ostriker J.P., 1990, in The Evolution of the Universe of Galaxies, ed
Kron R.G., Dordrecht, Reidel, p.25

\noindent Peterson B.A., Ellis R.S., Kibblewhite E.J., Bridgeland M.T., Hooley
T., Horne D., 1979, ApJ, 233, L109

\noindent Phillipps S., 1993a, in Observational Cosmology, eds Chincarini G. et
al, ASP Conference Series Vol. 51,

p.308

\noindent Phillipps S., 1993b, MNRAS, 263, 86

\noindent Phillipps S., Davies J.I., Disney M.J., 1990, MNRAS, 242, 235

\noindent Phillipps S., Disney M.J., 1986, MNRAS, 221, 1039

\noindent Phillipps S., Disney M.J., Kibblewhite E.J., Cawson M.G.M., 1987,
MNRAS, 229, 505

\noindent Phillipps S., Shanks T., 1987, MNRAS, 227, 115

\noindent Rocca-Volmerange B., Guiderdoni B., 1990, MNRAS, 247, 166

\noindent Roukema B.F., Yoshii Y., 1993, ApJ, 418, L1

\noindent Sandage A., Binggeli B., Tammann G.A., 1985, AJ, 90, 1759

\noindent Schade D., Ferguson H.C., 1994, MNRAS, 267, 889

\noindent Schechter P., 1976, ApJ, 203, 297

\noindent Schombert J.M., Bothun G.D., Schneider S.E., McGaugh S.S., 1992, AJ,
103, 1107

\noindent Smail I., Ellis R.S., Fitchett M.J., 1994, MNRAS, 270, 245

\noindent Tresse L., Hammer F., Le Fevre O., Proust D., 1993, A\&A, 277, 53

\noindent Turner J.A., Phillipps S., Davies J.I., Disney M.J., 1993, MNRAS,
261, 39

\noindent Tyson J.A., 1988, AJ, 96, 1

\noindent Tyson N.D., Scalo J.M., 1988, ApJ, 329, 618

\noindent van der Kruit P.C., 1987, A\&A, 173, 59

\noindent White S.D.M., Frenk C.S., 1991, ApJ, 379, 52

\noindent Wyse R.F., 1985, ApJ, 299, 593

\noindent Yoshii Y., 1993, ApJ, 403, 552

\newpage

\section*{Figure Captions}

{\bf Figure 1.} Model predictions for the basic model with $\alpha$(dwarf)
$= -1.5$, $\beta = 1$, $\Gamma = 1$ (see text for definitions).
Panel (a) shows the overall $z = 0$ LF
(solid curve) compared to the data points taken from LPEM. Dotted and dashed
curves show the separate contributions from giants and dwarfs. Panel (b) shows
the observed $B$ band number counts, as summarised in MSFJ and DPDMD (various
symbols) compared to the model predictions. Also shown is the contribution the
dwarfs would make if they were non evolving (long dashed curve). Panels (c),
(d) and (e) give the predicted redshift distributions compared to the data
for the magnitude ranges 20.5 to 21.5 (BES). 21.5 to 22.5 (CETH) and 23.0 to
24.0 (CSH). The number of missing redshifts is indicated by the size of the box
labelled "incompleteness".\\
\\
{\bf Figure 2.} As figure 1 but for $\beta=1$, $\Gamma = 2$.
\\
{\bf Figure 3.} As figure 1 but for $\beta = 1.5$, $\Gamma = 1$.\\
\\
{\bf Figure 4.} As figure 3 (ie. $\beta = 1.5$, $\Gamma=1$)
but with a cut-off to the evolution at $z = 0.7$.\\
\\
{\bf Figure 5.} As figure 2 (ie. $\beta = 1$, $\Gamma = 2$)
but with $\alpha = -1.75$.\\
\\
{\bf Figure 6.} As figure 2 (ie. $\beta = 1$, $\Gamma = 2$)
but with surface brightness selection effects taken
into account as described in the text.\\
\\
{\bf Figure 7.} The 'observed' LF at various redshifts for our best fitting
model (as per figure 6).

\end{document}